\documentclass[%
 reprint,
 superscriptaddress,
 showpacs,
 amsmath,amssymb,
 aps,
 prl,
]{revtex4-1}

\usepackage{siunitx}
\usepackage{color}

\usepackage{graphicx}
\usepackage{dcolumn}
\usepackage{bm}

\begin{document}

\title{Mapping a quantum walk by tuning the coupling coefficient}

\author{Ng Kian Fong}
 \affiliation{Centre for Quantum Technologies, National University of Singapore, Blk S15, 3 Science Drive 2, 117543 Singapore}
\author{Manuel J. L. F. Rodrigues}
 \affiliation{Centre for Advanced 2D Materials and Graphene Research Centre, National University of Singapore, 6 Science Drive 2, 117546 Singapore}
\author{Jos\'{e} Viana-Gomes}
 \affiliation{Centre for Advanced 2D Materials and Graphene Research Centre, National University of Singapore, 6 Science Drive 2, 117546 Singapore}
 \affiliation{Department of Physics, National University of Singapore, Blk S12, 2 Science Drive 3, 117551 Singapore}
\author{Alexander Ling}
 \affiliation{Centre for Quantum Technologies, National University of Singapore, Blk S15, 3 Science Drive 2, 117543 Singapore}
 \affiliation{Department of Physics, National University of Singapore, Blk S12, 2 Science Drive 3, 117551 Singapore}
 \author{James A. Grieve}
 \affiliation{Centre for Quantum Technologies, National University of Singapore, Blk S15, 3 Science Drive 2, 117543 Singapore}
 \email{james.grieve@nus.edu.sg}

\date{\today}


\begin{abstract}
We present a method to map the evolution of photonic random walks that is compatible with nonclassical input light. Our approach leverages a newly developed flexible waveguide platform to tune the jumping rate between spatial modes, allowing the observation of a range of evolution times in a chip of fixed length. In a proof-of-principle demonstration we reconstruct the evolution of photons through a uniform array of coupled waveguides by monitoring the end-face alone. This approach enables direct observation of mode occupancy at arbitrary resolution, extending the utility of photonic random walks for quantum simulations and related applications.
\end{abstract}

\maketitle

Quantum walks have been studied extensively in the context of quantum computing and simulation~\cite{Ambainis2008, Childs2009,Venegas-Andraca2012,Aspuru-Guzik2012}. Systems built around a quantum walk have been proposed as physical simulators for a wide variety of quantum~\cite{Kitagawa2010} and classical~\cite{Schreiber2012} phenomena. When random walks are leveraged for physical simulation the dynamics and evolution of the random walker as it traverses the graph are often of primary interest.

In the experimental domain, quantum walks have been demonstrated across a range of physical systems, employing trapped particles~\cite{Zahringer2010,Cote2006,Preiss2015} and photons~\cite{Broome2010,Chiodo2006,Peruzzo2010,Szameit2010,Schreiber2010,Corrielli2013}. The field of photonic quantum walks is particularly developed, owing to the ease of access to long coherence times and the ability to perform high fidelity manipulation of single particles using relatively low cost devices. Within this class, devices comprising waveguide arrays have proved popular due to their favourable scaling properties~\cite{Perets2008}.

When simulating a physical system, the evolution of the quantum walker is commonly inferred by monitoring fluorescence in the host device~\cite{Chiodo2006,Dreisow2012,Plotnik2014,Keil2015}. Unfortunately, this signal is able to capture only the intensity of the propagating optical modes, and so is unsuitable for following the evolution of more complicated inputs, for example multi-photon entangled states. Systems which leverage these inputs are only able to obtain a snapshot of the random walk at a fixed propagation length corresponding to the output plane~\cite{Matthews2013}. Even where the evolution of the walker is not of primary interest, access to this information may be desired as a diagnostic or calibration tool.

We have designed and implemented an optical circuit in which the evolution of a photonic quantum walk can be observed in a chip of constant length. In our system, a sequence of observations at the end face combined with appropriate tuning of the device parameters exposes the evolution of the input state in a manner that is compatible with single photon intensities, and extensible to multiphoton states. We demonstrate this technique by implementing the well-studied~\cite{Peruzzo2010} one dimensional, continuous-time random walk on a uniform graph.

\begin{figure*}[ht]
\centering
\includegraphics[width=0.8\linewidth]{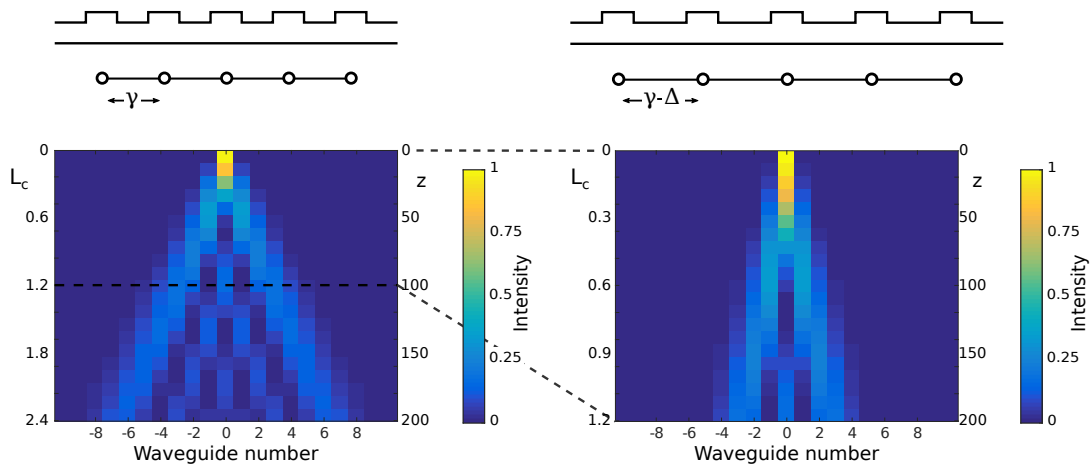}
\caption{The operational concept for a quantum walk in a waveguide array mapped via modification of the coupling coefficient (denoted $\gamma$). Top, a stylized profile of the waveguide array cut orthogonal to the propagation direction. Centre, the associated graph representation, a 1D array of vertices (circles) connected by edges (lines). Bottom, simulated evolution of a quantum walk on the structure depicted above, over an interval $\{z: 0,200\}$. The left column illustrates the initial array, for which $\gamma = 0.02$, while the right column illustrates a tuned configuration, with $\gamma = 0.01$. The associated range of $L_c$ is also shown, with the tuned configuration (right) exhibiting evolution over precisely half the range of its undistorted counterpart (left). All units are arbitrary.}
\label{fig:stretching-concept}
\end{figure*}

A continuous-time, discrete-space quantum walk on a one-dimensional graph comprising a set of vertices connected with edges (as depicted in Fig~\ref{fig:stretching-concept}, middle row) can be realized physically by injecting photons into an array of identical, continuously coupled waveguides. Following~\cite{Bromberg2009}, we describe the evolution of light in this structure using the Heisenberg equation for the creation operator $a^\dagger_k$. At a longitudinal distance $z$ along waveguide $k$, the creation operator is given by

\begin{equation}
\label{eqn:a-dagger}
a^\dagger_k(z) = e^{i \beta z}\sum_l{U_{k,l}(z)a^\dagger_l(z=0)},\quad U_{k,l}(z) = (e^{izC})_{k,l},
\end{equation}

\noindent where  $\beta$ is the propagation constant for the waveguides, $C_{k,l}$ are coefficients describing the coupling strength between adjacent waveguides and $izC_{k,l}$ encodes the coupling amplitudes at a distance $z$ along the array. The unitary transformation given by $U_{k,l}$ describes the evolution of input states across the array~\cite{Bromberg2009,DiGiuseppe2013}.

For a uniform array with only nearest neighbour coupling, $C_{k,l}=\gamma$ if $k,l$ are adjacent and $C_{k,l}=0$ for all other $k,l$. The evolution of the system depends on both $z$ and $\gamma$, which appear as a product in the elements of $\hat{U}$. It is common to parameterize such systems in units of ``coupling length'' $L_c$, defined as the value of $z$ for which $\gamma z = \pi/2$~\cite{Lifante2003a}. Two different implementations of a given system observed over the same interval of $L_c$ will evolve identically. Hence jumping rate and propagation distance determine the system dynamics on an equal footing, allowing them to be leveraged interchangeably in experiment.

A visual representation of this scheme is shown in Fig.~\ref{fig:stretching-concept}. The evolution of optical modes in a one dimensional quantum walk is obtained following Equation~\ref{eqn:a-dagger} and plotted for two different conditions. We plot the walk as a function of $L_c$ for each step of the evolution, enabling representation of the random walk on a standardized scale. In the example depicted, the second (tuned) state exhibits a 50\% reduction in the coupling coefficient, and consequently the evolution of its random walk is halted at half the range of its unmodified counterpart.\\

We propose to map a photonic quantum walk via the continuous tuning of the coupling coefficient of a uniform one-dimensional waveguide array. Unlike a previous implementation of this concept in which the wavelength sensitivity of the coupling coefficient was exploited to similar effect~\cite{Iwanow2005}, we tune the coupling coefficient directly by modifying the array itself. Experimentally, our approach requires a waveguide array be fabricated on a suitable platform to enable controlled modification of the coupling coefficient. In our earlier work~\cite{Grieve2017}, we developed a waveguide platform in the soft polymer polydimethylsiloxane (PDMS), and demonstrated continuous tuning of a single beamsplitter by stretching of the host chip. The ability to control the separation between neighbouring waveguides of an array through applied strain allows us to vary the coupling coefficient at a fixed wavelength, without significantly affecting the device length.

\begin{figure}[ht]
\centering
\includegraphics[width=\linewidth]{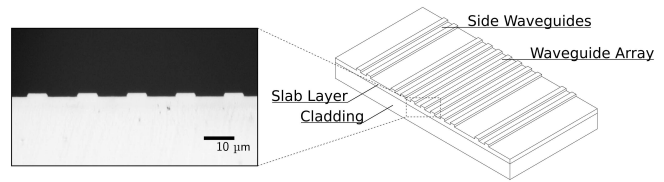}
\caption{A simplified representation of the array device, comprising an evanescently coupled array of rib waveguides with a \SI{17.5}{\micro\meter} pitch defined in a dual-layer polydimethylsiloxane chip. Also shown are the reference structures: pairs of waveguides flanking the main array. The optical microscope image shows the end-face of a typical device.}
\label{fig:array3d}
\end{figure}

As a proof of principle experiment, we fabricate a 1D array of waveguides on a PDMS chip using the casting technique previously reported~\cite{Grieve2017}. The device consists of an array of 51 single mode rib waveguides with a pitch of \SI{17.5}{\micro\meter}. On each side of the array, an isolated pair of waveguides act as ``reference structures'', with simple coupling behaviour~\cite{Lifante2003a} enabling independent measurement of the system-wide coupling coefficient. A simplified schematic of the chip is shown in Fig.~\ref{fig:array3d}, together with a microscope image of the device end-face. The chip is approximately \SI{7}{\milli\meter} long and supports polarization insensitive low-loss propagation (approx \SI{0.1}{\decibel/\milli\meter} at \SI{630}{\nano\meter}) over a wide wavelength range (\SIrange{450}{850}{\nano\meter}~\cite{Grieve2017}), suitable for use with single photon sources.

For controlled stretching along the direction transverse to mode propagation, we mount our chip on a jig driven by a miniature translation stage. Light is input into the array through edge-coupling a single mode optical fiber to the desired waveguide channel. The output of the device is then imaged onto a CMOS camera via a standard microscope objective. The output corresponding to each deformation of the device is recorded, then stacked and aligned to visualise the evolution of the quantum walk. The values of $L_c$ corresponding to each step in the walk are determined from the reference structures.

\begin{figure}[ht]
\centering
\includegraphics[width=0.9\linewidth]{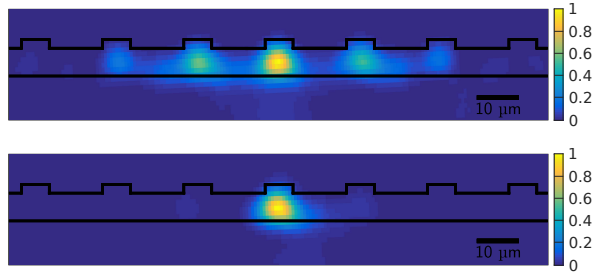}
\caption{Intensity distributions observed from the end-face of the waveguide array after a propagation distance of $\sim$\SI{7}{\milli\meter}, with the central waveguide coupled to a laser source of $\SI{532}{\nano\meter}$. The excited optical modes at the output are shown. Top: Optical modes observed from the initial device state. Bottom: Optical modes observed when the device is maximally stretched, with a chip-scale distortion of approximately +10\%. The decrease in the number of excited modes is a consequence of the decreased coupling coefficient between waveguides in the array, and enables us to probe the field structure at an earlier stage of the quantum walk. Measured values of $L_c$ are 0.13 and 0.38 for the stretched and unstretched states respectively.}
\label{fig:532slice}
\end{figure}

Intensity distributions at the end-face with no stretching and with maximum stretching are shown in Fig.~\ref{fig:532slice}. The spatial mode profiles are similar and appear minimally affected by the global distortion. This is supported by our observation that device tuning of this type results in a change of separation between neighbouring modes, but does not significantly modify the waveguide dimensions. A larger number of excited modes is observed for larger values of $L_c$. As the device is stretched, coupling between neighbouring waveguides decreases, resulting in a reduced power transfer from the source waveguide and hence fewer excited modes. The decrease in coupling due to stretching allows us to probe the quantum walk at different stages of the evolution.

The coupling coefficient between neighbouring modes is also expected to vary with wavelength, due to the wavelength dependence of the spatial mode profile. Several wavelengths of light can be used to map out distinct ranges of $L_c$. In this way it may be possible to extend the range of $L_c$ values obtained on a single chip by combining observations of several wavelengths.

\begin{figure*}[hbt!]
\centering
\includegraphics[width=\linewidth]{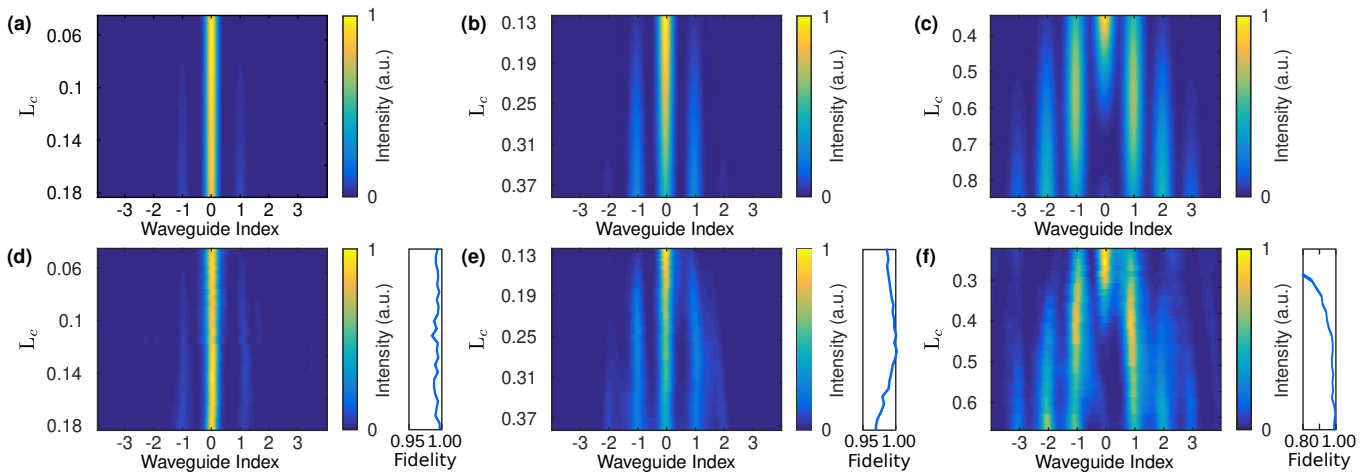}
\caption{Photonic quantum walks observed for a range of wavelengths: (a,d) \SI{450}{\nano\meter}, (b,e) \SI{532}{\nano\meter}, (c,f) \SI{630}{\nano\meter}. The experimentally observed quantum walks (d)-(f) were constructed from the output intensity distribution of the device (see Fig. \ref{fig:532slice}) observed over a large tuning range (up to $\sim$10\% of chip width). They are in good agreement with the simulated evolutions, (a)-(c) respectively, as confirmed by fidelities (included line plots, see Equation~\ref{eqn:fidelity}). The range ($L_c$) of the quantum walk changes with the wavelength of the input source, with the longest wavelength resulting in a larger range probed.}
\label{fig:combined}
\end{figure*}

Experimental and simulated quantum walks for wavelengths of \SI{450}{\nano\meter}, \SI{532}{\nano\meter} and \SI{630}{\nano\meter} are shown in Fig.~\ref{fig:combined}. Reconstructed evolutions qualitatively match their simulated counterparts for all three wavelengths. For \SI{630}{\nano\meter}, we observe an offset in the values of $L_c$ compared to the simulated distribution. At this wavelength, the reference structures did not provide sufficient resolution to capture the magnitude of $L_c$ with high precision. We are exploring alternative designs for reference structures in future experiments.

For quantitative comparison of the reconstructed evolutions, we compute the fidelity of observed spectra ($I_{expt}$) to the theoretically calculated values ($I_{theory}$), using the expression~\cite{Corrielli2013}

\begin{equation}
    \label{eqn:fidelity}
    F = \frac{\sum I_{expt} I_{theory}}{\sqrt{\sum I^2_{expt} I^2_{theory}}}.
\end{equation}

\noindent Fidelities for \SI{450}{\nano\meter} and \SI{532}{\nano\meter} measurements exceed 95\,\% for the full observed range, while for the \SI{630}{\nano\meter} data the fidelity falls below 90\,\% for the highly strained regions $L_c < 0.35$. We attribute some of this deviation to the previously discussed difficulty in determining $L_c$ values, but the presence of scattered light in waveguides $-3,3$ may also have played a role.

The results of our experiment demonstrate the validity of this method for observing the evolution of optical modes in quantum walks. By using a tunable photonic device to control the coupling coefficient between copropagating optical modes, we are able to reconstruct the evolution of quantum walks with extended range. Random walks are observed over a range of $L_c = 0$ to $0.8$, with wavelengths from $\SI{450}{\nano\meter}$ to $\SI{630}{\nano\meter}$. These measurements are performed on a low cost device, fabricated following polymer casting techniques that support rapid prototyping and are broadly accessible to the research community. Combined with a straightforward and intuitive mechanism to tune their optical properties, we believe chips of this type will be attractive to researchers in the field of photonic simulators.

This approach to observing a quantum walk provides a specific advantage compared to other chips used in quantum photonic experiments~\cite{Chiodo2006,Szameit2010,Dreisow2012,Plotnik2014} in that it does not require the use of fluorescence for visualisation of field evolution. This is an important consideration in experiments employing light of extremely low intensity, such as the single-photon regime. In particular, it should be possible using these devices to reconstruct the evolution of a wide variety of non-classical states in photonic random walks. For example multi-photon states can be used to simulate random walks in higher dimensional graphs~\cite{Peruzzo2010}, or of particles obeying non-Bosonic statistics~\cite{Sansoni2012,Matthews2013}.

The approach demonstrated here (utilizing chip-scale stretching) is directly applicable to any graph structure in which the jumping rate $\gamma$ may be factorized from the evolution operator. It can be extended to the study of non-uniform graphs by tailoring the strain response of the host chip (for example by varying the thickness of the substrate in the transverse direction). The introduction of out-of-plane distortions can be used to realize localized changes to the coupling coefficients. Using localized changes, the technique need not be restricted to continuous-time walks, as such tuning would also enable the effective removal of successive generations of splitters in a discrete-time, discrete-space system~\cite{Sansoni2012} to similar effect. Combined, we believe that these techniques have the potential to greatly expand the capabilities of the photonic quantum walk as a platform for quantum simulations.\\

\section*{Funding Information}

This work was supported by the Singapore Ministry of Education Academic Research Fund Tier 3 (Grant No. MOE2012-T3-1-009) and the National Research Foundation, Prime Minister’s Office, Singapore under its Research Centres of Excellence programme.

\bibliography{array}

\end{document}